\documentclass[twocolumn,showpacs,preprintnumbers,amsmath,amssymb,floatfix]{revtex4}

\usepackage{graphicx}
\usepackage{dcolumn}
\usepackage{bm}
\usepackage{amsmath,amssymb}
\usepackage{color}

\newcommand{\be}{\begin{equation}}
\newcommand{\ee}{\end{equation}}
\newcommand{\ba}{\begin{eqnarray}}
\newcommand{\ea}{\end{eqnarray}}

\begin{document}

\title{Left lane road electrification}

\author{Felipe J. Llanes-Estrada and Katja Waidelich}
\email{fllanes@fis.ucm.es}
\affiliation{%
Universidad Complutense de Madrid, Depto. F\'{\i}sica
Te\'orica I, 28040 Madrid, Spain.
}%

\date{\today}

\begin{abstract}
 We propose the left-side electrification of multilane motorways in 
suburban areas as a practical way of implementing on-road charging in order to reduce the weight and price of electrically-powered automobiles.  These could then be supplied energy en-route, significantly reducing the weight and cost of vehicle-mounted batteries. Our theoretical study is based on the circumstances of Madrid (Spain), a six million inhabitant region, but should be easily adaptable to other metropolitan areas.
\end{abstract}

\maketitle

\section{Motivation}

As is well known, transportation consumes a large fraction of the crude oil 
resources in the world, as reprinted in table~\ref{tabla:cuantotrans}.
\begin{table}[h] 
\caption{\label{tabla:cuantotrans}Percentage of oil consumption in transport, per year and geographic region, and projected growth. Source:~\cite{WEO}
 \vspace{0.1cm}}
\begin{center}
\begin{tabular}{|c|cccc|}\hline
Annum             &       1980    &  2000 & 2015 & 2030 \\ \hline
USA               &       52      & 66    & 71 & 72 \\ 
Western Europe    &       31      & 50    & 57 & 58 \\ 
Pacific           &       27      & 38    & 44 & 44 \\ 
Japan             &       23      & 36    & 39 & 38 \\ 
Eastern Europe/Eurasia &  24      & 37    & 45 & 45 \\ 
Asia (not OCDE)   &       25      & 36    & 42 & 51 \\ 
Middle East       &       32      & 36    & 39 & 43 \\ 
Africa            &       46      & 51    & 51 & 54 \\ 
Latin America     &       39      & 47    & 52 & 54 \\ 
{\bf World}       &       {\bf 38}      & {\bf 51}    & {\bf 54} & {\bf 57} \\ \hline
\end{tabular} \end{center}
\end{table}

The conclusion from the table is obvious. Half of the world's crude oil consumption is already due to transportation, and the trend (should no action be taken) is to increase further. 
Proven oil reserves can satisfy human demand at current levels for a good 40 years, but long term transportation solutions will have to be based on alternative energy sources. 

\subsection{Madrid's dependence on private commuting and import oil}

Our modern society is fully dependent on private automobiles for road transportation. Although the city of Madrid is endowed with an excellent city rail system, the regional railways are clearly insufficient and most of the commuting is, as in many other urban agglomerations, done on the roads.
Table~\ref{tabla:cuantoscoches} presents the number of utility vehicles, amounting to about half a percent of the world's total estimated at 650 million operating automobiles.

\begin{table}
\caption{Spain's and Madrid's automobile number circa 2010~\cite{DGT}. 
\label{tabla:cuantoscoches} \vspace{0.1cm}}
\begin{center}
\begin{tabular}{|ccc|}\hline 
Category & Madrid's region & Total in Spain \\ \hline
 & & \\
Family cars  &  3 376 000  & 22 145 000 \\
Trucks and Pick-ups & 669 000 & 5 192 000 \\
Buses & 11 300 &  62 000 \\
Motorcycles & 259 000 & 2 500 000 \\
Industry and agriculture  & 17 000 & 213 000 \\
Other motorized units & 78 000 & 855 000 \\
\hline
\end{tabular}
\end{center} 
\end{table}

80\% of homes in the Madrid region own an automobile, and 50\% of those not able to walk to work employ it daily, averaging to over two million vehicles daily on the road.

Madrid's energy dependence is absolute. The region produces barely 4\% of the energy it consumes. In the bigger national picture, Spain reaches 20.9\% of energy self-sufficiency, while a negligible 0.2\% of all oil consumed is domestically produced. Between 1985 and 2007 the consumption of Diesel fuels soared by a factor 4, while gasoline consumption remained stable.
The spanish economy as a whole pays foreign suppliers an estimated 33000 million euro annually for oil imports~\cite{cores} (other estimates elevate the tab to 39000M euros). 

Madrid hosts an eighth of the total automobile fleet of Spain, yielding up to 5000 million euro per year in the transportation sector. The total budget of the state's administration, for comparison, is a factor of two higher.
This wealth is handed out to its oil suppliers, the major ones being Nigeria, Lybia, Saudi Arabia, Iran, Irak, Mexico, Venezuela and Russia are the major ones. 

\subsection{Pollution by automobile fleet}

A diesel-powered vehicle typically emits to the atmosphere 150-200 grams of $CO_2$ per kilometer. The 22 million vehicles in Spain, assuming some 10000 km of yearly usage, are responsible for 33 million metric tons of carbon dioxide per year. This is about a fourth of the total emission (a figure also similar to the European total percentage). Best modern vehicles can lower the carbon emission to the level of 30-40 grams (from well to wheel) per kilometer, much improvement is possible, but there seems to exist a lowest floor for optimization. 

If the vehicle fleet could be made electric, a large part of the energy consumption could be turned to alternative sources. In the case of Spain, 35\% of the electricity production mix is not based on carbon sources, and this gain would be immediate. 

In addition, the rest of the energy would be produced at industrial installations outside the city (``pollute elsewhere'' principle)
which would drastically reduce the levels of Carbon monoxide (87\% of Madrid totals due to vehicle emissions) or Nitrogen oxides (66\%).

Sound pollution could also be reduced by adopting an electric vehicle fleet. For speeds above 50 km/h (city speed limit inside the inner belt of Madrid) there is little possible gain since most of the friction is produced in the tire-pavement contact~\cite{PhysToRasmussen} leading to 70-80 dB sound levels, but below that speed the explosion motor is the main source of noise. Electrical motors are remarkably silent, rated at about 10 dB.

\subsection{High cost of the electric vehicle}

By the end of 2010, the spanish government had hoped to reach 2000 electric vehicles and 540 static battery charging points. About two years later, the number of charging points has grown to 771~\cite{IDAE}, but the total number of circulating electric vehicles has not yet reached the original mark. By current sales data it appears that the goal will be met in 2013;
this is in spite of direct help to the purchase amounting up to 6000 euro per vehicle.
Looking further, China planed to have 50 000 electric cars in its territory by the end of 2012. This goal has not been met either. Israel's ``Better Place'' programme, a world referent (though also seeing a very slow start), scaled to Madrid's size, would entail 140 000 light electric vehicles and an unfeasible cost in the range of 400-800 million euro of public expenditure.
 It is clear  that the stumbling block to widespread adoption of the electric vehicle is its high cost at the dealer, and we believe that sales are not going to make the electric vehicle a mass-acquired product with this handicap.

\begin{widetext}
\begin{table}[h]
\caption{Estimated cost of several electrical vehicles upon reaching a consumer in Spain (not necessary official manufacturer prices) and several significant data about their battery's capacity and autonomy. Notes and abbreviations:
 Mitsubishi Innovative Electric Vehicle$^*$. Renault Zero Emission Fluence$^\dagger$. Includes range extender$^\&$. \vspace{0.1cm}
\label{tabla:coches}}
\begin{center}
\begin{tabular}{|c|cccccccc|}\hline
Model & MIEV$^*$ & Tesla Roadster & Nissan Leaf & RZE$ ^\dagger$ & Mercedes E-Smart & BYD e6 & Opel Ampera$^\&$ & Peugeot ION\\ \hline
Price (euro)   &  $>34000$    & $> 118\ 000$ &  33\ 400  &    28000       & $\simeq 40 000$     & 30 000 & 42 000 & $\simeq 30 000$ \\
Battery storage (kwh)&  16   & 51  & 24 & 22 & $\simeq 15$ & 60 & 16 & 16.5 \\
Range (km) & 150 & 340& 175 & 160 & 135  & 300 & 60 & 150 \\
Power (kw) & 49  & 40 & 80  & 70  & 30   & 75  & 111& 47  \\
Weight (kg)& 1100& 1238 & 1525 &1540 & 900 & 2205 & 1694 & 1110\\
\hline
\end{tabular}
\end{center}
\end{table}
\end{widetext}

Table~\ref{tabla:coches} presents estimates of current prices for electrical vehicles in the final consumer market.\\
These prices are well in excess of the typical 12000 euro or less that a consumer pays for a small family utility vehicle conventionally powered by a diesel engine. Fiscal incentives and subventions to purchase electric cars, of order 2000 to 6000 euros, can hardly overcome the price step. While some reduction of the price is achievable with mass-production and sales in the future, one should expect electric cars to still remain a luxury product for a while as we now argue.

Battery prices at the industrial end are currently estimated to be up to 1000 euro/kwh~\cite{bosch}. A two-battery set with 70 kwh energy storage and 180 km range can cost somewhat less, some 15 000 euro; to achieve a price reduction of 10 000 euro per vehicle to curb it into marketability, it is necessary to significantly reduce the cost of its electric batteries. This is not foreseeable by any price reduction of Lithium, other metals, or manufacturing costs~\cite{Argonne1}.
 Mass production by 2020 is expected to lower prices at most to the 200-300 euro/kwh, but not lower, since about 30\% of the production cost seems to be independent of production volume. 

Looking again at table~\ref{tabla:coches} we note that the typical range is 7-8 km/kwh. Therefore, achieving a 300 km range with an electric vehicle will still imply a battery cost, even after mass production is achieved and the price per kwh is down to 300 euro/kwh, of 9000 euro, which is the entire price tag of a gasoline fueled vehicle of similar characteristics.
New ideas are clearly a necessity if the benefits of electric automotion are to be realized.

The advantages of the electric vehicle, once purchased, are significant in terms of cheaper energy consumption, lower maintenance and operating costs, lower pollution, etc. But the public, generally not holding large amounts of capital to spend upfront, prefers a cheaper initial deal at the expense of the higher maintenance and operating costs of diesel fueled vehicles.

Can one find a scheme where the upfront purchase is significantly cheapened, even if later the expenses are higher? 
We concur with others in the concept of on-road charging and contend that providing the electricity on the road via a conductive delivery system~\footnote{Unlike inductive systems that in our view is not as demonstrated a technology yet.}, in analogy with the railways, is a possible answer. Batteries would then store much less energy and make the car cheaper. 

In this article we discuss this possibility at a theoretical level and present preliminary investigations that suggest that the option should be further considered.

\section{The fleet's energy supply}
\subsection{Reasonable consumption}

Simple estimates of well-to-wheel efficiencies~\cite{Tesla}
for very efficient fossile fuel vehicles (models similar to the Honda Civic) yield about 2.2 kilometers per kwh of gasoline energy. Electric vehicles with proven efficiencies (such as the Tesla Roadster) can reach 4.8 km/kwh, a good factor of 2 above their fossile fueled counterparts. Low-weight hybrid vehicles fare somewhere between the two estimates. However a lot of the energy expense associated to an electric vehicle is at the power plant and distribution. The actual range per kwh in the battery is 9.1 km, which is a key figure for us. \\
If these nine kilometers are run at a moderate highway speed of 90 km/h,
the power consumption is 10 kw. This figure is for optimal vehicle circulation. 

In table \ref{tabla:potencias} we recapitulate some estimates from table~\ref{tabla:coches} of the power needed by actual vehicles reaching the consumer market.
\begin{table}[h]
\begin{center}
\caption{\label{tabla:potencias} Power in kw of several electric vehicles available in the market in Spain.
\vspace{0.1cm}}
\begin{tabular}{|c|c|}
\hline
MIEV               & 47 \\
Nissan Leaf        & 80 \\
Renault ZE Fluence & 70 \\
Mercedes E-Smart   & 30 \\
BYD e6             & 75 \\
\hline
\end{tabular}
\end{center}
\end{table}

Let us now check these maker's estimate with simple physical arguments and see how much more power should we wish delivered to a vehicle.

Aerodynamic resistance to the vehicle's advance can be estimated as
\begin{equation}
F =  \frac{C_d}{2} \rho v^2 S
\end{equation} 
where the air density is about $\rho\simeq 1.22$ kg/m$^3$, and the effective area perpendicular to the advance (area times the aerodynamic coefficient) is, for a small family car such as the Opel Astra, $C_d S\simeq 0.68$ m$^2$.
Altogether, at the reference speed of 90 km/h, this implies a power consumption of $P=6.5$ kw.

Ground friction can be estimated as
\begin{equation}
F_r = C_r \times (mg)
\end{equation}
which, for a rolling motion friction coefficient $C_r=0.03$ makes a vehicle loaded to 1500 kg of mass spend an additional $11\ 250$ watt. 
Assuming a 70 \% efficiency in the electric motor and the transmission,
we obtain an estimate for the power consumption in optimal rolling of
$P=25$ kw, for a car which is analogous to convenience vehicles in current use.

Finally one needs to plan for the additional stress caused on the battery upon acceleration, necessary upon joining the highway, but also upon conditions of dense traffic when braking and reacceleration become necessary.

A 1500 kg automobile accelerating in 10 s from rest to 90 km/h will need some $46.9$ kw. \\
Let us therefore adopt the figure of 50 kw as the desirable maximum power to be provided to an average small automobile by an energy delivery system.

 Can one exceed this demand by sustained periodic traffic circumstances?
We can negatively answer with a simple model with a cosinusoidal profile
\begin{equation}
v(t)= v_0 \cos^2\left(\frac{2\pi t}{T}\right) \ .
\end{equation}
The root-mean squared average acceleration is (after a simple integration over a half-period)
$$
\bar{a}= \frac{\sqrt{2}\pi v_0}{T} \ .
$$
To exceed 50 kw one needs, for the 1500 kg vehicle, with 40 meter races between two full stops, and peak velocity of 47 km/h, acceleration and breaking times of about 6 seconds. We do not see this intensity of traffic as realistically sustainable in a motorway and therefore conclude that 50 kw can be taken as a standard in most circumstances.
If a vehicle instantaneously needs more power delivered (air condition and on-board amenities are further expenses), it can detract it from the batteries; if in smaller need, it can reload them. All in all, we think that 50 kw is a reasonable demand on the external road energy supplier.

\subsection{Electricity supply}

With a horizontal area seldom exceeding 5-6 m$^2$, it is clear that vehicle mounted photovoltaic panels are not an option for family vehicles in motion, since only 1-2 kw of energy can be obtained. Hydrogen fuel cells are a more practical option for producing energy on board the vehicle by chemical means, but they are pricy.

The current bet in many countries, including Spain, is for the deployment of electric vehicles that do not produce their own power but instead store it in batteries. Supply will be provided by both charging points at home or parking lots, and battery pack replacement stations, ``electrolineras'' in spanish language, in analogy with currently used ``gasolineras'', following the battery renting model.
We find this model unpractical if scaled to a large fraction of the current automobile fleet, given the enormous  inmobilized capital invested in batteries that the owner of the electric supply station needs to have at customer's disposal.

What we propose is to have an electric line running parallel to the highway, such that the driver can hook to the same and obtain the standard 50 kw of power from the line with wich to operate the vehicle and recharge its batteries without stopping. 
Thanks to such traction power feeding system, middle and long range travel become possible with very small battery packs. \\
This concept seems a reasonable option and is the model employed by developed railway lines. Many cities also operated electric buses, trolleys, in the past (some do to this day such as Lyon, Genoa or Ghent) supplied by aerial electric lines.  
Its adoption would imply the possibility of much reducing battery packs in electric vehicles.

In the particular case of Madrid, the municipalities most distant from a highway susceptible of electrification are the (sparsely populated) towns of Rascafr\'{\i}a (26 km to A1 near Buitrago), Valdemaqueda (21 km to M-501 near Navas) and Valle San Juan (20 km to A3 near Villarejo). Commuters from these towns would need batteries with a range of 50 km at least. Nevertheless, for most of the six million inhabitants it would be sufficient to have batteries with ranges of the order 20-30 km for all regional driving purposes, once highway electrification had been carried out.

\section{Railway electric lines}\label{railways}

The railway sector has the relevant experience for long-distance electric supply
of vehicles~\cite{Siemens,board}.
Key developments to cheapening electric locomotive units and grid deployment in the past decades where the AC-rectifier locomotive, the thyristor propulsion control for smoother acceleration, vacuum circuit breakers for 50 kVolt operation that allowed 65 km intervals between substations, and so on for a long list of advancements. All this technology base should be relatively off the shelf for road electrification, if adopted.

Estimates of the cost of electrification of a km of railway have increased from some 100 000 dollar/km through the 1970's~\cite{board} to order one million euros per kilometer nowadays. We have consulted the budgets of several such civil works in Spain and conclude that a budget capped at 1 Million euro per kilometer is sensible,
and even more, contemporary enterprises entail much challenging demands due to the high speeds achieved by the train units. This will not be the case for personal electric vehicles.
Of the total cost, about 15-20\% should be allotted to the electric substations feeding the system. Another 5\% should be reserved for the extension of the transmission lines of the utility company. The maintenance of the system should consume yearly about 1\% of the initial capital cost.

An important issue that needs to be kept in mind is the necessity of balancing loads on utility generators at power plants, in terms of phase and voltage, but this we find more worrysome for locomotives than for individual cars, since the load difference in the act of connecting to or disconnecting from the grid is much smaller for the small utility vehicles.

Railway electrifications usually have three options to deliver power: overhead contact lines, third rail installations, or an overhead conducting rail. We now focuse on the overhead contact line, closest in spirit to our proposal.

For up to 100 km/h the catenary can be of the type ``catenaire economique'' with the contact line suspended from several droppers near the poles, to offset the wire slope near a support point, but without a second wire running above it to guarantee very homogenous suspension height through the entire line~\cite{little}. It is of course true that a second, suspension wire and a reduced span length between poles make a much more straight contact line, thus reducing sag and consequently arcing and wearing of the wire, though this can also be achieved by increased tension in the wire. Technical specifications can for example be found in~\cite{ergon}.

The material of choice for the contact cable is electrolytic copper or copper alloys (with e.g. magnesium), because upon oxidization it forms a few micrometer film of CuO, CuO$_2$, that has still good conducting properties.
The wires are grooved so they can be clamped on the side opposite to where the vehicle's pantograph will establish contact. Typical forces exerted on  them by the pantograph at moderate speeds are the weight equivalent of 10 kg. More precisely~\cite{zoller} the pantograph SBS 65 used for tests exerted a static force of 70 N on the line, with this figure increasing with speed up to 200 N. In a two locomotive train, the back pantograph, though exerting the same force on average, had to tolerte larger oscillations (up to 215 N instead of 165 N for the leading pantograph) at high speed.  The tolerance for the uplift of the lines is of order 10 cm in railways.

\section{Left lane electric line}

Railways are often supplied by a suspended catenary line placed directly on top of the rails. This is possible by the large standardization of rail wagons, since the operators are one or few companies per rail line and acommodate their vehicles to the height of the infrastructure. We think this is a less favorable option for a road electrification given the many different vehicle heights. If large trucks were to circulate by the same electrified lane ever, small family cars would be forced to carry extremely long poles to contact the electricity carrier. 

To have the electricity line on the right side of the road, where it would naturally fit in countries with wheel on the vehicle's left by the smaller vehicle speeds on the right lane, has the inconvenience of numerous line interruptions due to the highway exits on that side.

Many underground railways and airport cars take electricity from the surface railing. This we do not deem safe enough in the open highway, given the possibility of people, animals, or machines shortcircuiting a system on the ground.

Therefore we opt for temptatively proposing an electrification of the left lane of the road, with two suspended cables on the side. 

\subsection{Electric power necessary}

At a speed of 90 km/h the maximum occupation of the lane that should 
be allowed is 25 vehicles (implying a very short breaking distance of 40 meters). At 50 kw of standard supply per vehicle, this gives us an estimate of 1.25 Mw supplied per kilometer of line. If one lane for each of the two traffic directions is electrified, the total power needed would be 2.5 Mw/km.

A standard 1 Gw Westinghouse nuclear reactor can therefore supply 400 km of thus partially electrified highway. 
The main network of highways in the region of Madrid had 632 km as of mid-2010 (out of a 2610 km network that includes secondary roads). Therefore we would conceive this road electrification to be supplied by one additional power plant with two reactors or equivalent generation capacity by other means.\\
This power generation (and more) will be necessary in any case if the electric vehicle is adopted as a standard. The only difference with our proposal is that the energy be transfered on the road instead of static charging points.

\subsection{Voltage choice}

The choice of a specific voltage is very much region-dependent and worldwide carmakers will need to be flexible in adapting transformers to their vehicles to feed typical electrical motors with 100 V of DC.

For a road electrification it seems natural to adopt the local railway standards, to easily find civil contractors with the necessary expertise. 
Turning to Madrid, the underground metro system (5 km characteristic distance) is electrified at 600 V of DC. Most regional railways (50 km ranges), to reduce Ohmic losses, are fed at the higher 3000 V
 DC. Finally, high speed railways (with 500 km long lines) built in the last two decades operate at 25 kV AC.

Let us briefly examine Ohmic losses. Combining Joule's and Ohm's laws, the handed-out power and the lost power are related by means of the voltage and the electrical resistance of the line
\begin{equation} \label{PJoule}
P_{\rm Joule} = \frac{R P^2}{V^2} \ .
\end{equation}
For a wire of uniform cross section, the resistance is elementarily given by
$  R = \rho \frac{L}{S} $. 

Steel at a temperature of 18  Celsius has a resistivity coefficient
$0.098\times 10^{-6}$  Ohm m, and Aluminum at equal temperature $0.028\times 10^{-6}$  Ohm m~\cite{gerthsen}.
A cable of section 500 mm$^2$, yields a resistance per kilometer of 0.2 and 0.05 Ohm respectively.\\
Often used is the so called ``Condor'' cable, a composite of both materials, with a resistance of about 0.07 Ohm/km, somewhat intermediate. 
These are good materials for the feeding wires. But the contact cable will have to be manufactured from copper, as discussed above in section~\ref{railways} for the railway case.

If losses are to be kept below 1\%, there is a maximum distance along which one wants to transport a given amount of power, obtainable from Eq.~\ref{PJoule} as
\begin{equation} \label{maxL}
L\leq \sqrt{ 1\% \frac{V^2}{(R/km)(P/km)} } \ .
\end{equation}
Considering the three railway voltages mentioned above, we find that, if we want to guarantee the supply of 1.25 Mw to each kilometer of cable with resistance 0.07 Ohm/km, the approximate maximum lengths $L$ are
\begin{itemize}
\item For 600 V DC, 200 m.
\item For 3000 V DC, 1000 m.
\item For 25 kV AC, 8.4 km.
\end{itemize} 

The final choice would have to be made by car designers in cooperation with the utility company. However, in view of these numbers, an interesting possibility is to combine an underground high-voltage cable (at 25 kV or up) connected at kilometer intervals to a transformer and rectifier that step down the voltage to 3000 V DC and feed the aerial voltage-carrying line. This arrangement guarantees acceptable losses, and the intensity of current carried even at maximum conceivable load (cars fed 50 kwatt at 10 m intervals with 3000 volts) is 1666 A, well within the capability of railway copper contact wires. 

The last step of current transformation occurs in the vehicle where an automobile-mounted passive DC to DC converter (usually a solid state device) should output the typical 100 V DC required by the motor. 

\begin{widetext}
\begin{figure}[h]
    \includegraphics[width=13cm,angle=-90]{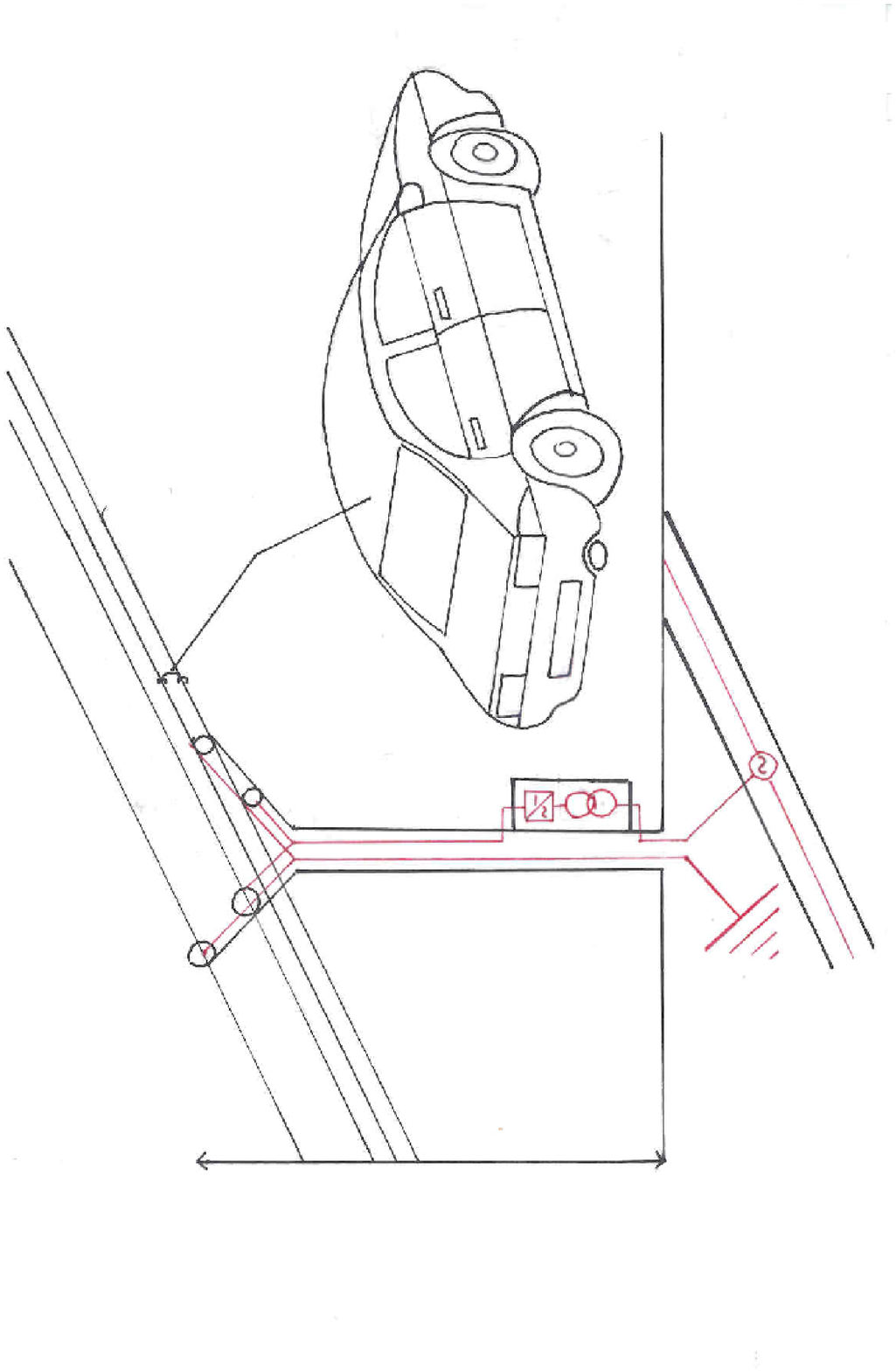}
     \caption{\label{fig:hookedcar}
Sketch of the electric line feeding a rolling automobile. An underground AC cable feeds an aerial 3000 volt DC line, a third cable provides the ground voltage.
A $Y$-disposition allows simultaneous feeding two lanes in opposite directions of the highway where this be possible. Otherwise the system supplies only the left lane in one direction.
Automobiles should connect to the cable by a side pantograph-antenna.
            } 
\end{figure}

\end{widetext}

The system would share elements with a common road illumination setup, with buried cables and a vertical steel pole every 25-50 meters. Except, more like in railway electrifications, two catenaries (one grounded and one carrying the 3000 V voltage) would hang from them. Every kilometer or so the underground cable would feed the signal-carrying aerial.
The situation is sketched in Fig.~\ref{fig:hookedcar}.

\subsection{Electricity grid}

Urban areas typically already count on good electricity networks. The Madrid region is supplied with 220 and 400 kV aerial lines that can be found near every major highway.
Of course, the generalization of electric transport would require a reinforcement of these electric grids, which requires much specialized research.

Highway side electric service stations for electric cars can serve a dual role. Currently they are conceived as containing battery-changing points and fast-charge outlets at about 800 V. \\
One could easily conceive them also as hosting transformers to receive very high voltage from the power grid and feeding it, once stepped down, to the road electrification at intervals of order 10 km (again sensible in densely populated areas).

\subsection{Automobile connector}

Railways are connected to their top electricity supply by a pantograph.
These are articulated, bulky arms: typical weights are in the order of 100 kg (for example 109 kg for the DSA-350S design) that are too encumbering for automobiles.
Sideways connection for automobiles could be effected by a pole, not much unlike a side-mounted radio antenna of somewhat larger length and thickness than usually found in cars. This automobile pantograph should be articulated
with two knees because automobiles are not much guided by a rail (see later in section~\ref{habits}).
To reduce the risk of electricity shock, we would recommend a coaxial line with the inside cable carrying the voltage and the outside grounded. Many dielectrics can maintain the necessary electric field differences as seen in table~\ref{tabla:dielectricos}. The choice of a dielectric would in the end be made by mecanical considerations, durability and price.
\begin{table}[h]
\caption{\label{tabla:dielectricos} Breakup voltage for several dielectric materials~\cite{dielectricos} easily allowing a 1 cm thick coaxial insulation.}
\begin{center}
\begin{tabular}{|cc|} \hline
Material & Breakup field (Volt/cm) \\
Polyester & 340 000 \\
Polyamid & 360 000 \\
Aramid & 50 000 \\
Epoxy   & 24 000 \\ \hline
\end{tabular}
\end{center}
\end{table}

\begin{figure}[h]
    \includegraphics[width=7cm]{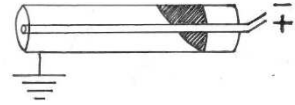}
    \includegraphics[width=4cm]{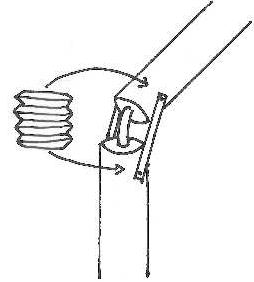}
    \includegraphics[width=6cm]{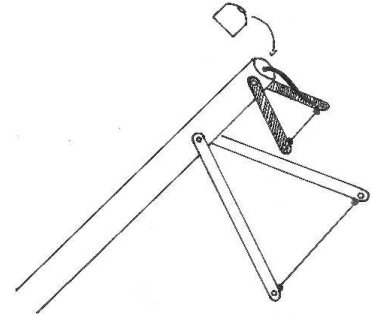}
     \caption{\label{fig:coaxial}
Sketch of the coaxial antenna, which is somewhat thicker than usual radio antennae, supporting the  current feeding the vehicle. The outside steel cylinder is grounded by the second aerial thereby providing ground voltage for the vehicle's mass for safety reasons as well. It can also be coated to avoid corrosion and other damage. The inner wire carries the 3000 V potential and is protected by an insulating polymer (colored black). The antenna needs to be articulated to allow for slight distance changes between automobile and aerial. At the contact end dampers should be provided to reduce noise, and the contact wires should be engineered to reduce arcing upon contact.} 
\end{figure}

A sketch of the antenna's section is given in figure~\ref{fig:coaxial}.
We propose that the articulation should allow for a tolerance of half a meter in the sideways distance to the automobile. Larger separations should mean that the driver wanted to separate from the connection.
Common practice in electrified railways~\cite{Siemens} allows for a tolerance to pantograph-electric line oscillation of 10 cm in perpendicular to the pantograph. Accepting this figure and allowing the half-meter margin to the driver means that the pantograph has to be able to correct the distance by 40 cm, so its design must necessarily be articulated. This should pose no inconvenient to automobile development companies.

The contact to the line is performed by collector strips or wires. In railway usage these are made of carbon to reduce wear of the feeding electric line by friction (an alternative to explore is a cylinder rolling over the line instead of sliding). In any case they need to have low contact electrical resistance, good conductivity, high melting point, and good mechanical properties.

\section{Driving habits} \label{habits}

We would recommend a speed limitation to 90 km/h or similar in the electrified left lane, to reduce mechanical stresses and simplify the design of the electrical line and the pantograph hooking to it from the vehicle. This is not overconstraining in many highways of metropolitan areas, with three or more lanes, where faster traffic can be carried by the middle lane.

In fact, in many states, for example California, the left lane is sometimes already reserved for vehicles with high passenger number, not for the fastest vehicles.

In the case of Madrid, the average speed in its beltlines has varied between 50 and 65 km/h in the period 2004-2008, so that 90 km/h would not be a real restriction except at night hours.

\begin{figure}[h]
 \begin{center}
    \includegraphics[width=7cm]{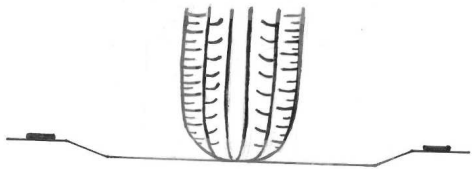}
     \caption{{\label{fig:rueda}}
A small depression in the pavement with side visual and sound bands 
helps the driver keep the vehicle at constant distance from the electric line.}  
      \end{center}
\end{figure}

A greater difficulty lies in the driver maintaining the vehicle aligned with the electric line to avoid loosing contact. 
We recommend for this purpose a small depression in the left end of the road of about 2 cm depth and 1 m width (to allow for a 50 cm standard variation of mean distance between vehicle and electric line). This substitute of a railway's track offers no real resistance to a driver wishing to abandon it, but with proper visual and pavement sound aid, it is effective to help him stay connected to the aerial. The situation is depicted in Fig.~\ref{fig:rueda}.
In addition, one could adopt trolley-bus native devices such as the so 
called ``contact shoes'' that assist in avoiding dewiring in the absence of a rail guidance.

In a full lane, with many car units hooked to the line, the wire is pushed back by the leading automobiles and the trailing ones will presumably have a harder time making contact, taking larger wire-vehicle distance oscillations. The tension of the wire $T$ (achieved by counterweights) has to be large enough to avoid the formation of transverse waves that may dewire some vehicles. The formula 
\be
T > \left(\frac{v({\rm km/h})}{3.6}\right)^2 \lambda 
\ee
gives the minimum tension in terms of the linear density of the wire $\lambda$ in kg/m and the car velocity, about 90 km/h in our design. A factor 1.4 between the two sides of the inequality is recommended by railway experts~\cite{Siemens}.

\section{Financial feasibility}
\subsection{General considerations}
We would like to examine the question of whether the electric car economic model can closely mimic the current oil-based model in terms of financing, with an equal upfront expense upon purchase and similar energy costs later. This can work assuming that the manufacture price for both is quite the same,  which can be achieved once freed from expensive battery packs.

Since there is no direct experience with such electrifications, we will take as orientation the cost of electrifying a railway line or illuminating a road.
We have exposed in section~\ref{railways} that the electrification cost for a railroad is estimated at a million euro per kilometer~\footnote{we have also consulted other civil undertakings of analogous nature in Spain, such as the illumination of highways, and see no reason to rebutt this figure.} as amply allowing for the electrification we propose today.
We will call this variable capital cost per unit length, $C/L$ to keep the discussion general. The cost of electrical off-road grid infrastructure is accounted for in the electricity price and need not concern us any further.

By current oil prices, a commuter in Spain is paying about 9 euro cents per kilometer. If her vehicle would be moved by electricity, her energy cost would instead be about 4.2 cents per kilometer. Thus, the bare energy price is cheaper for electricity by more than four cents per kilometer. This price difference will be denoted by $\frac{\delta p}{L}$.

What we want to argue next is that  these excess four cents per kilometer can finance the needed on-road electric infrastructure. The state acts only as a regulator and the private exploitation of the electrification is possible. This will be feasible with certain minimum traffic levels which we establish next.

We will, for an order of magnitude estimate, neglect the very much fluctuating weekend and vacation traffic, and concentrate on the $N_d=235$ yearly working days where commuter traffic is quite constant, leaving the rest as surplus payments.

It is clear that near the city, denser traffic makes the electrification more profitable. Let the variable $x$ run from $0$ at the city center through positive values of the distance to it along the highway.
The number of cars at any one point of the highway will be the variable $n(x)$, with $\frac{dn(x)}{dx}$ being a continuous approximation to the actual number of cars entering and leaving the way at the exits.
The number of vehicles that have adopted electricity as an energy means can be represented by multiplying with a ratio smaller than 1,  $r(x)$ or $r(t)$ (depending on whether one wishes to stress or not its time dependence).
As a function of time one expects $r(t)$ to be a population-logistic curve, with initial exponential growth and later saturation.

\subsection{Loan repayment}
Let the infrastructure capital $C(t)$ be financed by a state-guaranteed loan, and  assume for the sake of the discussion a yearly interest rate of 
$TAE=5\% $.
To avoid increasing the debt, the yearly repayment must be 
$P(t) \geq TAE\times C(t)$.
The revised owed capital will therefore be, at year $t+1$,
\be \label{sucesion}
C(t+1) = C(t)- \left( P(t) - TAE \times C(t) \right) .
\ee
If the time for repayment is chosen to be $\tau$, so that $C(\tau)=0$,
and solving Eq.~\ref{sucesion} for a constant payment $P(t)=P$, we find 
\begin{equation}
P = \frac{C(0)\times \left(1+TAE\right)^\tau}{\sum_{i=0}^{\tau-1}\left(1+TAE \right)^i } \ .
\end{equation}
Taking $\tau=30$ yr (a generation's span), the payment will be about
$P/L=65\ 050$ euro/annum.
This is the typical revenue that each kilometer of highway needs to yield for the investment to break even.

This cost needs to be shared among the users. To keep the total user payment at or below the current cost of operating with fuel-based cars, we obtain the condition
\begin{equation}
\frac{P}{N_d n(x)} < \delta p \ ,
\end{equation}
from which the minimum number of cars needed daily on the road falls out as
\begin{equation}
n(x) \geq \frac{P}{N_D \delta p} \ .
\end{equation}
For the reader's ease we have collected the resulting number of cars as a function of the energy source price difference in table~\ref{tabla:numerocoches}.
\begin{table}[h]
\caption{Number of circulating electric cars necessary on each work day at kilometer $x$ of a motorway's electrified lane, assuming a yearly interest rate on capital investment of $TAE=5\%$ and an electrification cost of $1M$ euro per kilometer, as function of the price difference between fossile fuel and electricity needed to drive that kilometer.
 \label{tabla:numerocoches}\vspace{0.3cm}}
\begin{center}
\begin{tabular}{|c|c|}
\hline
$n(x)$ (car number)  & $\delta_p$ (oil-less electricity \\ & in cents/km) \\ \hline
$13\ 840$ & 2 \\ 
$9\ 226$ & 3 \\
$6\ 920$ & 4 \\
$3\ 460$ & 8 \\
  \hline
\end{tabular}
\end{center}
\end{table}
All of these numbers are very far from saturation; note that with a speed of $90$ km/h and car separation of 50 m, the maximum number of cars that a highway lane operating twelve hours in a given day can take is 21 600, well above those figures. 

\subsection{Case study: A6 in Madrid}\label{A6}

A6 is one of eight major radial roads and gives commuter access to the city from the northwest. We here consider a stretch of about 50 km inside the region of Madrid.

The Mean Daily Traffic Intensity was measured in 2002 to be 71 438 incoming vehicles and 78 872 outgoing vehicles at kilometer $x=19$ from the center, 
where the road had four useful lanes in each direction of circulation.

Rounding off to 70 000 daily commutes, and with $\delta P\simeq 4$ cents/km (that is the current situation), the electrification of the left lane would become profitable for $r\ge 10\%$, that is,  with one in ten vehicles being electric.

The regional planner would face the question of when and how far to provide the electrification. One can sensibly think that vehicles within $10$ km of the city would opt for commuting on their batteries and not paying the electrification fee, recharging at night instead. Therefore, if the number of vehicles that make the electrification profitable is calculated to be $N$, it must be that 
\begin{eqnarray} \label{cuandoelec}
\left\{
\begin{tabular}{cc}
${\rm if}\  x>10\ km \ $& ${\rm then}\  N< r (t+2yr)\times n(x) $\\
${\rm if}\  x<10\ km \ $& ${\rm then}\  N< r (t+2yr)\times n(10)$  \ .
\end{tabular}
\right.
\end{eqnarray}
Recall that $r$ is the fraction of electric vehicles, which we evaluate at retarded time by an estimated 2 years (typical time span between decision and operativity of a civil work of these characteristics).

\begin{figure}[h]
 \begin{center}
 \vspace{1.cm}
    \includegraphics[width=8cm]{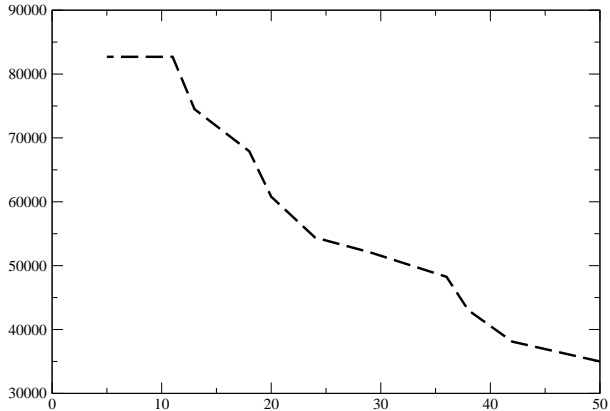}
     \caption{\label{fig:traficoA6}  Estimate of Mean Daily Value of car number on A6 as function of the distance to the center, saturating at kilometer $x=10$ from the center.}
     \end{center}
\end{figure}

We do not have at hand traffic data for all points of A6, so we have made an estimate with the known data (traffic at city accesses, at the furthest point of A6 where two tunnels go under the mountain chain in the limit of the Madrid region, and others) and filled the missing traffic data by making it proportional to the population that potentially accesses that highway (the census is known) at each kilometric point $x$. The outcome is shown in Fig.~\ref{fig:traficoA6}.

From these data for $n(x)$ and our prior financial discussion, it follows that it is profitable to electrify the entire 50 km length of the highway when the percentage of electric vehicles reaches 20\%.

To decide on the electrification of a given kilometer all that is needed is to know the rate $r(t)$  at which the electric vehicle will be adopted (if it finally is).
The initial growth phase of a technology revolution is exponential, but we need to know the characteristic time.

\begin{widetext}
It might be of help to consider the adoption of cellular telephones, the latest wide-reaching technology revolution. The characteristic time of its logistic curve happens to be about $\tau=2$ yr.

Coming to the electric car, we consider $\tau=8$  months through $\tau= 3$ yr in table~\ref{tab:crecimiento}.

\begin{table}[h]
\caption{\label{tab:crecimiento}
Number of electric vehicles estimated for the next years according to the unknown value of  $\tau$ (time in which the number is multiplied by $e\simeq 2.718$). The adoption of cellular phones was characterized by $\tau=2$ yr. The initial condition is an estimate of 2000 vehicles in 2013 (this was Spain's goal for 2010 that is being reached with delay). For each $\tau$ value, the left column is the percentage of the fleet and the right number the actual number of fully electric automobiles. \vspace{0.3cm}}
\begin{center}
\begin{tabular}{|c|cc|cc|cc|cc|}\hline
 Year& $\tau =$ & 8 month & $\tau =$ & 1 yr &$\tau =$ & 2 yr & $\tau =$ & 3 yr \\
\hline
 2013 & 0.0091 & 2 000   & 0.0091 & 2 000  & 0.0091 & 2 000 & 0.0091 & 2 000\\
 2014 & 0.041  & 8 960   & 0.025  & 5 440  & 0.015  & 3 300 & 0.013  & 2 790\\
 2015 & 0.183  & 40 180  & 0.067  & 14 780 & 0.025  & 5 440 & 0.018  & 3 900\\
 2016 & 0.82   & 180 120 & 0.18   & 40 170 & 0.041  & 8 960 & 0.025  & 5 440\\
 2017 & 3.7    & 807 340 & 0.50   & 109 200& 0.067  & 14 780& 0.034  & 7 590\\
 2018 & 16     & 3 618 800& 1.3   & 296 830& 0.11   & 24 360& 0.048  &10 590\\ \hline
\end{tabular}
\end{center}
\end{table}

It is seen that the electrification will not be profitable in the four-year horizon of an administration.
\end{widetext}
\subsection{Net Present Value}
Looking farther ahead, one may still ask when is the electrification expected to be profitable. A useful analysis tool is the Net Present Value, defined as
\be
VNP = \sum_0^{N} \frac{C_n}{(1+d)^n}\ .
\ee
Here $C_n$ are future cash inflows in year $n$ due to the infrastructure, and $d$ is the discount rate (whose meaning is that future income could be obtained by alternative means, for example buying bonds, and therefore the gain $C_n$ very far in the future is less competitive). 
If the Net Present Value turns positive, the infrastructure should be undertaken, as it will be more profitable than alternative investments. However the value of $d$ is usually uncertain and depends on market conditions.

\begin{figure}[h]
\vspace{0.5cm}
 \begin{center}
    \includegraphics[width=7cm]{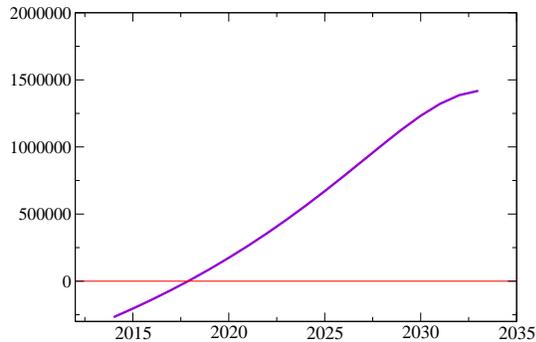}
     \caption{\label{fig:VNP2}   Net present value of the electrification of one kilometer of A6 in Madrid, in Euro, as a function of the year of construction, employing $\tau=2$ yr for the characteristic time of adoption of the electric vehicle
and a discount rate of 6\%.}
      \end{center}
\end{figure}

We plot in figure~\ref{fig:VNP2} the resulting Net Present Value for the electrification of A6 following from the analysis of subsection~\ref{A6}.
From the figure one concludes that, although if electrification construction undertaken in 2017 would still incur losses the first few years, it would break even in the 30 yr span of the capital repayment. If the electric vehicle is indeed adopted, construction undertaken \emph{after that date} is profitable in the long-term.

\subsection{User payment}

If the electrification is to be self-sustainable, as we have argued it can be in metropolitan areas, and exemplified with the A6 highway in the Madrid region, charge of the service has to be made to the user.

This would be a great application of Power Line Communication~\cite{PLC}, that is already being deployed for electricity fees incurred by vehicles at a stationary charging point. The on-board electronics can easily pick up a modulated high frequency signal through the electric network (again, the electric service stations spaced every ten kilometers could host the corporation's computer billing the charge).

Other ways of charging drivers for their use of the electric line more in line with current toll collection (such as a radiowave device, or credit card charge) are also possible but less attractive.

\section{Conclusions and outlook}

To summarize the article, we have proposed to supply electricity to electric cars on highways by means of an aerial line suspended at about 3 m height on the left side of the left-most lane (in right-circulating countries such as continental Europe or the USA) . \\
Reasonable parameters to planify the infrastructure include a provision of 50 kw per vehicle, taking into account that consumption of a small family car in ordinary circumstances will be in the range of 25-35 kw at a reference speed of 90 km/h. \\
A possible voltage to be employed is 3000 volts in direct current, with an auxiliary high voltage line underground, feeding the first at kilometer intervals.\\
There is no obvious technological roadblock to the deployment of such systems, although much engineering work and standardization would have to be undertaken.

We have shown the profitability of the electrification under reasonable assumptions on financial markets, if the electric car is indeed adopted.
As a gross reference, one should like to have 7 000 electric cars employing the electrified lane in each working day.\\
As the user is concerned, his transportation costs would closely parallel those of the current gas-based model, with a smaller upfront investment in the vehicle purchase than current electric cars require, but a higher operating cost.

Many questions remain and grant further research, but many are of a social nature and common to other methods of supplying electric cars. One for example is whether the commuters would be limited to regional travel with their low storage-capacity vehicle if the road electrification was not extended. One should like to carry out a similar analysis to interstate or national-level road systems taking into account the data of larger areas with population spread in several nodes. 
In the case of Spain, a length of order 8 000 km of road should be electrified to have a national impact (we estimate the cost of this deployment as equivalent to a fifth of the current profit of the entire electricity sector over a decade).

It is also interesting to compare the analysis with the railway electrification. The last plan undertaken in Spain for conventional railways dates back to 1979 (high-speed railways are automatically electric-powered)~\cite{informeICAI}. As of Dec. 31st, 2008, 60\% of the conventional rail network  was completely electrified, with 5350 km missing. The total electrification of the system was encouraged and claimed to be profitable already in 1945~\cite{Bueno}. \\
Based on train circulation numbers and power consumed per unit we estimate the total power in the high-speed line Madrid-Seville to be of order 140 MW for a 500 km stretch of railway. The consumed power is a fourth of the requirement for our proposed road electrification. The installed power in that railway line, with 500 Mwatt installed, is right on. Feeding stations  (supporting two train traction units of 8.8 Mwatt each) are separated 50 km because of the use of 25 kV high voltage: our road electrification requires feeding more often however.  These estimates show that, while technologically challenging, what we suggest is not far from currently existing infrastructure. And the technology is proven in transport (railways, tramways, trolleys...) unlike the new ideas of magnetic resonator couplings~\cite{Yu} for wireless transfer that are very promising, but would take much longer time to reach deployment.

Since transport is responsible for the consumption of 50 \% of crude oil, any electrification achieved is a way to reduce the world's dependence on this resource. 
We hope that, given the importance of things that are at stake, our contribution at a theoretical level warrants further specialized investigation.

\begin{acknowledgments}
  An older spanish-language monograph from which this work is abstracted can be found at our University's webpage under \\
{\tt http://teorica.fis.ucm.es/~ft11/\\ 
LEFTLANE.DIR/coche.pdf} \\
 All information that the scholar needs from that work has been collected in this english version, but a student may find the additional (if lengthy) discussion of use. 
\end{acknowledgments}

\newpage


\begin{thebibliography}{50}

\bibitem{WEO} World Energy Oulook 2008, report of the International Energy Agency.


\bibitem{DGT} According to statistics of Spain's Traffic Direction,
\\
{\tt www.dgt.es/portal/es/seguridad\_vial/estadistica \\ /parque\_vehiculos/por\_provincia\_y\_tipo\_parque/}. \\
We have rounded off to thousands of vehicles given data volatility, excepting the smaller bus number.

\bibitem{cores} 2012 data from the ``Corporaci\'on de Reservas Estrat\'egicas de Productos Petrol\'{\i}feros'', spanish government, see
{\tt www.cores.es} .

\bibitem{EnergiaMinisterio} ``La Energ\'{\i}a en Espa\~na 2007''. Report of the Ministery of Industry, Energy and Trade, Secretary of Energy.
Official General Publication Catalog, {\tt http://www.060.es}.
 

\bibitem{PhysToRasmussen}  R.~O.~Rasmussen and P.~R.~Donavan, Physics Today {\bf 62}, Dec. 66 (2009).


\bibitem{IDAE} Data from the governmental office {\tt{www.idae.es}}, the ``Instituto para la diversificaci\'on y ahorro de la energ\'{\i}a'', 2012. 

\bibitem{bosch} J. Olalla (Bosch Spain), communication at ``Madrid Ecocity'' workshop,  Madrid, 12$^{\rm th}$ April 2012.

\bibitem{Argonne1} ``Costs of Lithium-Ion Batteries for Vehicles'', Report ANL/ESD-42 of the Transport Research Center of Argonne's National Laboratory.


\bibitem{Tesla}    M. Eberhard and M. Tarpenning``The 21 Century Electric Car'', Tesla Motors Inc., Julio 2006, \\
{\tt www.veva.bc.ca/wtw/Tesla\_20060719.pdf} \ .


\bibitem{board} Transportation Research Board of the National Academy of Sciences, Special Report 180, Washington, 1977.

\bibitem{little} See article by E.G. Scharm and A.D. Little, ``Capital \& Maintenance costs for fixed Railroad Electrification Facilities'' in ref.~\cite{board}, page 42.


\bibitem{ergon} J. Brooks,
``Network Lines Standard Guidelines for overhead line design'', Ergon Energy Co, 
{\tt{http://www.ergon.com.au/}}


\bibitem{zoller} H.~Z\"oller, ``Development of pantographs for German Railways traction vehicles'', Elektriche Bahnen {\bf 49}, 168-175 (1978).

\bibitem{gerthsen}  C.~Gerthsen and D.~Meschede, Physik,
Springer, Berlin;  22. revised ed. (2005).


\bibitem{dielectricos} Parlex Corp.,  ``Flexible circuit dielectric base material options'', see {\tt http://www.parlex.com}.


\bibitem{barea} P. Barea L\'opez and O. Mart\'{\i}nez \'Alvaro, ``Distribuci\'on del tr\'afico en los grandes corredores espa\~noles'', 2003 study.

\bibitem{Siemens}
F. Kiessling, R. Puschmann and A. Schmieder,
``Contact lines for electric railways'', Publicis Corporate Publishing, Munich \& Erlangen, 2001;
A. Steinel, ``Elektrische Triebfahrzeuge und ihre Energie Versorgung'', Oldenbourg Industrieverlag, Munich, 2004.

\bibitem{PLC} Niovl Pavlidou {\it et al},  ``Power Line Communications: State of the Art and Future Trends'', IEEE Communications Magazine, April 2003.

\bibitem{informeICAI} A. Garc\'{\i}a \'Alvarez and M.P. Mart\'i{\i}n Ca\~nizares, Anales de Mec\'anica y Electricidad, May-June issue, 2009. 

\bibitem{Bueno} P. Gonz\'alez Bueno, Revista de Obras P\'ublicas, Nov. 1945, page 515.

\bibitem{Yu}
X. Yu {\it et al.}, ``Wireless energy transfer with the
presence of metallic planes'', App. Phys. Lett. {\bf 99}, 214102, 2011.

\end{thebibliography}
\end{document}